\documentclass[conference]{IEEEtran}
\IEEEoverridecommandlockouts
\usepackage{balance}
\usepackage{graphicx}
\usepackage{multicol}
\newcommand{\fix}[2]{{\bf FIX}\footnote{{\bf #1:} #2 }}
\renewcommand{\fix}[2]{}

\begin{document}

\title{Privacy, Security and Trust\\in the Internet of Neurons}


\author{\IEEEauthorblockN{Diego Sempreboni}
\IEEEauthorblockA{\textit{Department of Informatics} \\
\textit{King's College London}, UK \\
diego.sempreboni@kcl.ac.uk}
\and
\IEEEauthorblockN{Luca Vigan\`{o}}
\IEEEauthorblockA{\textit{Department of Informatics} \\
\textit{King's College London}, UK \\
luca.vigano@kcl.ac.uk}
}

\maketitle

\begin{abstract}
Arpanet, Internet, Internet of Services, Internet of Things, Internet of Skills. What next? We conjecture that in 15-20 years from now we will have the Internet of Neurons, a new Internet paradigm in which humans will be able to connect bi-directionally to the net using only their brain.
The Internet of Neurons will 
provide new, tremendous opportunities thanks to constant access to unlimited information.
It will empower all those outside of the technical industry, actually it will empower all human beings, to access and use technological products and services as everybody will be able to connect, even without possessing a laptop, a tablet or a smartphone. The Internet of Neurons will thus ultimately complete the currently still immature democratization of knowledge and technology. 
But it will also bring along several enormous challenges, especially concerning security (as well as privacy and trust). 

In this paper we speculate on the worldwide deployment of the Internet of Neurons by 2038 and brainstorm about its disruptive impact, discussing the main technological (and neurological)
breakthroughs required to enable it, the new opportunities it provides and the security challenges it raises. We also elaborate on the novel system models, threat models and security properties that are required to reason about privacy, security and trust in the Internet of Neurons.
\end{abstract}

\begin{IEEEkeywords}
New Internet paradigm; Brainwaves; Human computer; Security; Privacy; Trust
\end{IEEEkeywords}

\section{Introduction: from the human computer to... the human computer}
We all carry around a computer, regardless of who we are, how old we are, where we live, what job we do, what education we received. No, we are not talking about your laptop, your tablet or your smartphone. We are talking about your \emph{brain}.\fix{Luca}{Rispetto alle promesse del rebuttal di NSPW mancano i punti 2, 8, 10 e 11, almeno in parte}

In fact, the term ``computer'' has been in use from the early $17^\mathrm{th}$ century, way before electronic computers became available. It was introduced simply to mean ``one who computes'', namely a person whose job is to perform complex mathematical calculations. 
In that sense, people often speak of ``human computer'' to make this distinction clear.\footnote{In his famous 1950 paper~\cite{Turing1950}, Alan Turing wrote: ``The human computer is supposed to be following fixed rules; he has no authority to deviate from them in any detail.''} Throughout the centuries, human computers, working alone or 
in teams, have provided significant contributions to groundbreaking scientific discoveries, ranging from trigonometry to astronomy, to the  dawn of nuclear energy and nuclear weapons (e.g., the complex computations crucially related to nuclear fission in the Manhattan Project) and to the space race~\cite{Shetterly16}.

When electronic computers became available in the second half of the $20^\mathrm{th}$ century, human computers became useless, and ``human computer'' is nowadays mainly used to refer to individuals with prodigious powers of mental arithmetic who display their abilities in theaters or TV shows. Electronic computers also brought along a revolution that has transformed the economic, social, educational, and political landscape in a profound and indelible manner: the \emph{net}.


The technical foundations of the Internet were laid by the Advanced Research Projects Agency Network \emph{ARPANET}~\cite{ARPANET} towards the end of the 1960s.
Soon after, new overseas nodes of the network were created and the definition of the standard TCP/IP officially launched the \emph{Internet} as a set of interconnected networks through these packet switching protocols. 

Advances in hardware and software at the end of the $20^\mathrm{th}$ century enabled mobile connectivity to billions of laptops and (smart)phones. This \emph{Mobile Internet} gave rise to the \emph{Internet of Services (IoS)}~\cite{schroth2007web,mandula2015mobile}, with the flourishing of e-commerce, health-care portals, booking services, streaming websites and, last but not least, social networks. This redefined entire segments of the economy in the first decade of the $21^\mathrm{st}$ century, and was soon followed by the \emph{Internet of Things (IoT)}, a network of physical devices, vehicles, home appliances and other items embedded with electronics, software, sensors, actuators, and connectivity which enables these objects to connect and exchange data~\cite{gubbi2013internet,Al-Fuqaha2015,osisanwo2015internet,mandula2015mobile}. 




The next, and $5^\mathrm{th}$, evolution of the Internet is expected to be the \emph{Tactile Internet}, which has been defined by the International Telecommunication Union (ITU) as a network that is based on 5G and combines ultra-low latency with extremely high availability, reliability and security~\cite{Simsek2016,TactileInternet}. 
The Tactile Internet will encompass human-to-machine and machine-to-machine interaction, enabling tactile and haptic sensations and the control of the IoT in real time. It will unleash the full potential of the fourth industrial revolution (a.k.a.~Industry 4.0), and revolutionize the way we learn and work through the \emph{Internet of Skills} (a.k.a.~Human 4.0,~\cite{Dohler2017}). Although 5G has long passed the embryonic stage, and the testing phase is now underway, extra works must to be done to cover 5G security challenges~\cite{ahmad2018overview,schneider2015towards} in order to consider 5G a fully adoptable technology. However, capitalizing on 5G and ultra-low delay networking as well as on AI and robotics, the Internet of Skills will enable the real-time delivery of skills in digital form remotely and globally.


After this brief overview of the past, and the near future, of the Internet, it is time to ask what will came next. 
We conjecture the return of the human computer, but in a different guise. We predict the coming of the next, and maybe ultimate, phase of the Internet evolution: 
the \emph{Internet of Neurons} will rest upon a novel paradigm in which humans are able to connect bi-directionally to the net using only their brain. The Internet of Neurons will provide new, tremendous opportunities thanks to constant access to unlimited information. It will empower all those outside of the technical industry, actually it will empower all human beings, to access and use technological products and services as everybody will be able to connect, even without possessing a laptop, a tablet or a smartphone. The Internet of Neurons will thus ultimately complete the currently still immature democratization of knowledge and technology. But it will also bring along several enormous challenges, especially concerning security (as well as privacy and trust).

In the rest of this paper we speculate on the worldwide deployment of the Internet of Neurons by 2038 and brainstorm about its disruptive impact, discussing the technological (and neurological)
breakthroughs required to enable it, the new opportunities it provides and the security challenges it raises. We also elaborate on the novel system models, threat models and security properties that are required to reason about privacy, security and trust in the Internet of Neurons.

We proceed as follows. In Section~\ref{sec:IoN}, we introduce the Internet of Neurons. In Section~\ref{sec:PST}, we discuss privacy, security and trust issues in the Internet of Neurons. In Section~\ref{sec:concl}, we draw conclusions.

\section{The Internet of Neurons: From brainwaves to packets, and vice versa}
\label{sec:IoN}

In a 2014 interview~\cite{HawkingUSAToday}, Stephen Hawking said
\begin{quote}\em
We are all now connected by the Internet, like neurons in a giant brain.
\end{quote}
Although Hawking is famous for his predictions (as well as for his scientific results, of course), in this case he was not prophesying the advent of what we call the Internet of Neurons. However, it is interesting
to note that he used the same keywords (we found this quote when we googled ``Internet of Neurons'' to see if somebody had already had the idea) and that, in a brain, like in the Internet, it is actually all a matter of connectivity.

How would connectivity work in the Internet of Neurons? At the root of all our thoughts, emotions and behaviors is the communication between neurons within our brains. \emph{Brainwaves} are produced by synchronized electrical pulses from masses of neurons communicating with each other. Hence, to realize the \emph{brain-net}, which is one of the frontiers of \emph{brain-computer interaction} and thus of human-computer interaction,
we need to interface brainwaves with the packets that are received and sent by computers or other external devices.\footnote{Note that we are here assuming that the ``normal'' network will still be operating through packets, although by then advances in quantum computing (i.e., computing using quantum-mechanical phenomena, such as superposition and entanglement) might have provided for new modes of data transmission. But this is a topic for another paper.}

Some approaches have already been proposed, and prototypical devices and software built, for the realization of 
\emph{brain-computer interfaces}~\cite{Graimann}. We can summarize the methodology behind brain-computer interaction, through a brain-computer interface, as the following sequence of steps:
\begin{enumerate}
\item Collect brainwaves by recording activity directly from the brain (invasively or non-invasively) in real-time.
\item Convert the complex waveforms of brainwaves into data.
\item Encode the parsed information and issue action instructions.
\item Feed back the externally perceived information in real-time in the form of signals that the brain can read (possibly through a stimulating device).
\end{enumerate}
Note that the system must rely on intentional control, i.e., users must choose to perform a mental task whenever they want to accomplish a goal with the brain-computer interface.

Nowadays, it is already possible to detect and process brainwaves (e.g., using EEG sensors placed on the scalp) and a number of solutions have been proposed to provide a form of uni-directional communication and thus address at least steps 1) and 2) of this methodology. Let us consider 
three interesting examples. 
The neurotechnology company ``Neuralink'' was founded in 2016 by Elon Musk and others with the aim of developing an ultra-high-bandwidth implantable brain-computer interface to connect humans and computers~\cite{winkler2017elon}. 
While Neuralink is still in early stages, the ``Brainternet'' project~\cite{linesbrain} has developed an apparently more rudimentary but effective technology that streams brainwaves onto the Internet (by converting brainwaves into signals and streaming them to an online server using a Rasperry Pi computer). 
In 2018, the startup Neurable will release the VR game ``Awakening'' in which the gamer's brain essentially acts as mouse thanks to a brain-scanning headband paired with software that interprets the neural signals, thus allowing for hands-free control~\cite{Strickland}. Other application areas that brain-computer interfaces are currently being developed for are, for instance, education (e.g., for monitoring of students' attention in real time) and 
medical care (e.g., for monitoring and treatment of Parkinson's 
and other serious brain diseases, with the eventual goal of human enhancement as aspired  by Neuralink and other projects).

These technologies are promising, but they are still far from addressing steps 3) and 4) in a satisfactory way. The Internet of Neurons will require more than a uni-directio\-nal information flow; it will require a bi-directional information flow, in which 
\begin{itemize}
\item brainwaves are translated into data and 
\item data is translated into signals that the brain can parse.  
\end{itemize}

Some exploratory research is being carried out that attempts to bridge neuroscience with computer science and telecommunications, but  brain-computer bi-directional information flow is still largely unchartered territory. 

\begin{figure}[t]
\includegraphics[scale=0.075]{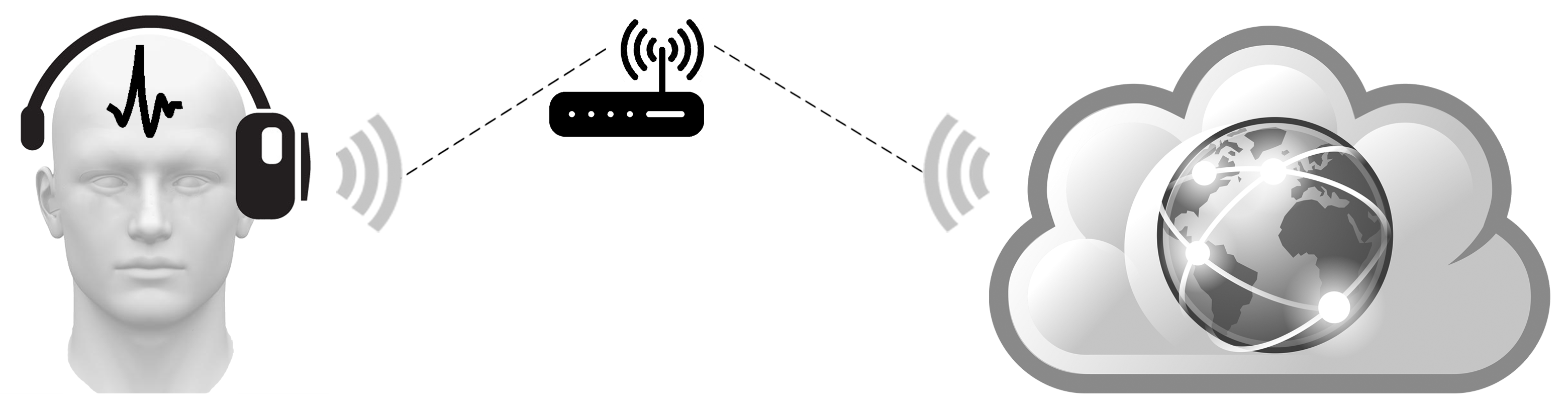}
\caption{Configuration 1 --- Bi-directional brain-Internet connection by means of a wearable device}
\label{fig:pod-notrust}
\end{figure}

Nonetheless, we conjecture that by 2023, in five years from now\footnote{Actually, it is not really important whether it will be in 10, 20 or 30 years, but rather that this will happen for sure, in one form or the other. And this time we should do it right, considering security from the start, unlike what happened when Internet was first designed as pointed out Danny Hills in~\cite{LoAndBehold}:
\begin{quote}
\emph{Because the internet was designed for a community that trusted each other, it didn't have a lot of protections in it. We didn't worry about spying on each other, for example. We didn't worry about somebody sending out spam, or bad emails, or viruses, because such a person would have been banned from the community.} 
\end{quote}
}, advances in neurology and in brain-computer interaction, combined with technological innovations, will have led to the creation of a device able to connect the human brain to the Internet bi-directionally, and without resorting to any invasive surgical operations. This device won't be bulky; it will be portable, light and chargeable inductively so that we will be able to connect to the Internet anywhere anytime. It could take the form of a lightweight headphone like in Fig.~\ref{fig:pod-notrust}, or more likely simply be a button-like pod that we will attach to our temples. Or it could even be a tiny implant, although non-invasive procedures are typically to be preferred. 

The device will communicate bidirectionally with the brain via brainwaves (as illustrated by the brainwave symbol on the forehead of the human in Fig.~\ref{fig:pod-notrust}) and with the Internet via wireless communication (as illustrated by the standard symbol) to and from appropriate routers. The device must thus be capable of reading the brainwaves in real-time, more or less like EEG readers are capable of doing now, but it must also be capable of interpreting the brainwaves and transform them into their digital version, sending the coded version to the Internet. The device must also be capable of receiving incoming data, convert it into brainwaves (Step (3)) and send them to the brain (Step (4)).

Being able to convert data into brainwaves and vice versa is necessary in this phase. Progress in Machine Learning, AI and Big Data have made it possible to interpret brainwaves~\cite{neuralencoding} mapping them with words or pictures creating a valid and applicable brainwaves-to-digital and digital-to-brainwaves codification. 
Feeding back the converted data into the brain requires techniques capable of stimulating the brain with signals. \emph{Electroconvulsive therapy (ECT)}, \emph{rapid transcranial magnetic stimulation (rTMS)} and \emph{magnetic seizure therapy} are 
techniques able to deliver stimulation pulses through the tissue directly to the brain, even wirelessly~\cite{chen2015wireless,grossman2017noninvasive}.

\begin{figure}[t]
\includegraphics[scale=0.075]{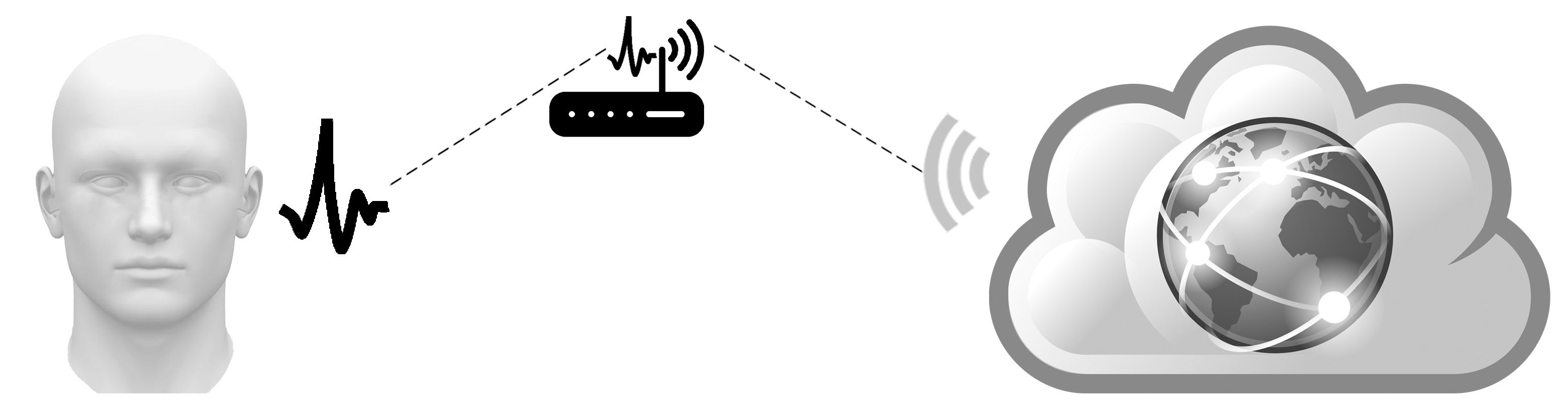}
\caption{Configuration 2 --- Bi-directional brain-Internet direct connection} 
\label{fig:podless-notrust}
\end{figure}

We also conjecture that advances in software and hardware will make sure that in 2038 there will be no more need for any wearable device to connect: as depicted in Fig.~\ref{fig:podless-notrust}, humans will be able to connect to the Internet directly with their brainwaves, possibly through routers that ``read'' brainwaves remotely (say from a distance of a few meters like wireless routers do now with wireless signals), and transform the brainwaves into data and vice versa (i.e., brains downloading and uploading information from the network). This may sound like the killing argument of ``tin-foil-hat conspiracy theorists'', who wear hats made from one or more sheets of aluminum foil in the belief that the hat will shield the brain from threats such as electromagnetic fields, mind control, and mind reading. However, this too is not really science fiction: research is ongoing on developing sensors that can be used to monitor the human electroencephalogram without electrical or physical contact with the body~\cite{Harland02,Prance08,Rendon17}. There is still a long way to go until these sensors are actually able to do more than just monitor but actually allow for the full realization of the four steps that we described above, but several foundation stones for the Internet of Neurons have been, or are in the process of being, laid so it is necessary that we start thinking about the privacy, security and trust challenges that will plague the Internet of Neurons. Some of these challenges will mirror the challenges that are plaguing Internet as we know it today, but other challenges will be novel and even more intriguing.


\section{Privacy, security and trust}
\label{sec:PST}

The potential offered by the technological revolution underlying the Internet of Neurons will be as varied as the problems related to privacy, security and trust that it will cause. In order to reason about these problems, we will need to provide suitable definitions, where a security definition is typically provided by combining a system model with a threat model and with one or more security properties that the system should guarantee even in the presence of an attacker. 
In the following, we discuss the main features of such models and properties for the Internet of Neurons. 
In our analysis, we thus take into account the two configurations suggested in the previous section, where the connection is made with or without a device, pointing out analogies with, and differences from, current research and technologies. 




\subsection{System model}

To provide a model of the system means to give a clear, and preferably 
formal, definition that provides enough detail to be able to understand and specify how the system behaves, encompassing both when it behaves correctly and securely, and when it behaves in unexpected and insecure ways. 

In the  security literature, security models have been formulated in a number of different ways. For instance, encryption and decryption operators are typically described by means of mathematical formulas along with some algebraic structure to capture the operators' properties; 
security protocols are typically described by means of state transition systems that specify how the knowledge of the protocol agents evolves over time; 
firewalls are typically described by means of sets of rules regulating how packets are filtered;
access control systems are typically described by means of security policies, requests and permissions;
software systems are typically described directly by their source code (or by the specification that can be learned or inferred by interacting with the code) or by dataflow and/or control flow specifications. These are just some examples, but all of them have in common the need to represent the infrastructure and how information flows among the system's agents (a.k.a.~principals or entities). 



For example, 
for Configuration 1 (Fig.~\ref{fig:pod-notrust}), we can identify the following agents:
\begin{itemize}
\item the human being,
\item the device,
\item the router(s),
\item the Internet,
\end{itemize}
connected by the following communication channels:
\begin{itemize}
\item a short-range channel between human being and device,
\item a medium-range channel between device and router,
\item a long-range (and possibly wired) channel between router and Internet.
\end{itemize}
Different protocols will be used to transmit information over these channels. The channel between the device and the router and the channel between the router and the Internet might actually employ protocols similar to the wireless protocols that we are already using today ---  in fact, if we are interested in a formal analysis of the system, we could even abstract away the channel between the router and the Internet and simply consider a medium--to-long-range channel between device and Internet. The channel between the human being and the device will, however, require new protocols able to translate between brainwaves and data packets, as the technologies that we discussed in the previous section are attempting to do.

For Configuration 2 (Fig.~\ref{fig:podless-notrust}), we can identify the following agents:
\begin{itemize}
\item the human being,
\item the router(s),
\item the Internet,
\end{itemize}
connected by the following communication channels:
\begin{itemize}
\item a medium-range channel between human being and router,
\item a long-range (and possibly wired) channel between router and Internet.
\end{itemize}
As before, different protocols will be used to transmit information over these channels. 
We expect that it will be possible to generalize to this configuration the protocols developed for the short-range brain-device communication in Configuration~1.

In both cases, the model of the configuration will need to be extended with models of the agents (including their actions and their states), of the security protocols used (including routing protocols), of the messages being sent, of the cryptography used and so on. We expect that many of the modeling languages and techniques that are in use today will be applicable with reasonable extensions, except of course for the translation brainwave-data, which will require considerable work. A starting point could be the formalization of this translation as a new cryptographic operator that encodes brainwaves into data along with the inverse operator that decodes data into brainwaves; identifying and formalizing the properties of these operators won't be easy though.

\subsection{Threat model}

A number of questions need to be answered in order to provide a threat model:
\begin{itemize}
\item \emph{Who is the attacker?} Is he an outsider or an insider? Is he an agent (a human or a machine) trying to attack the communication between the human and the Internet? Is he perhaps the router, or even the human itself? What if the human behaves honestly but makes mistakes, or thinks ``wrong thoughts'' (whatever they may be) that make the system vulnerable? How would social engineering look like in this case?

\begin{figure}[!t]
\includegraphics[scale=0.075]{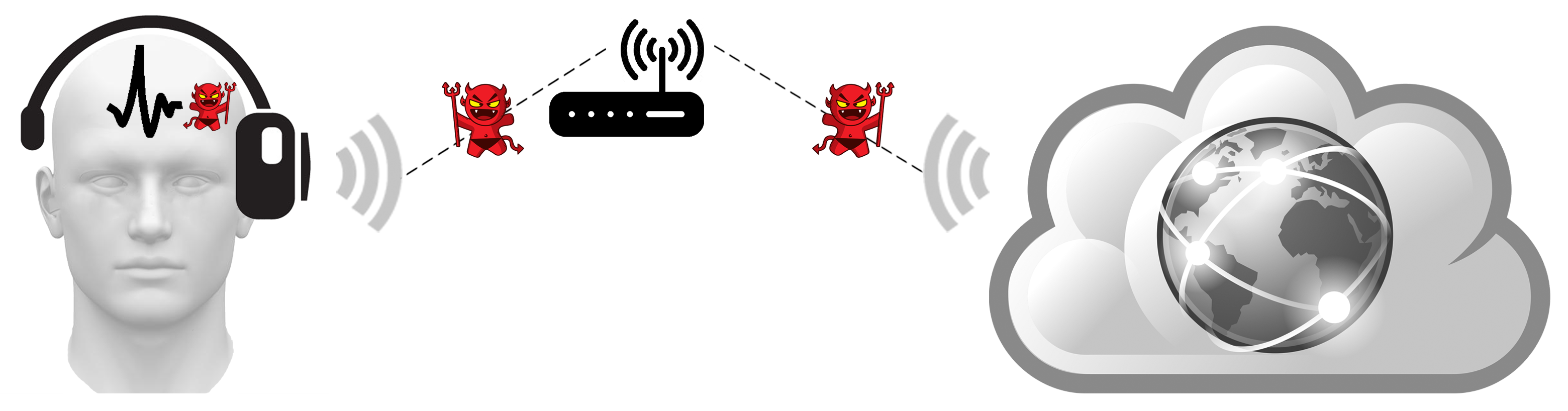}
\caption{Possible attacker locations in Configuration 1}
\label{fig:threatdevice}
\end{figure}

\begin{figure}[!t]
\includegraphics[scale=0.075]{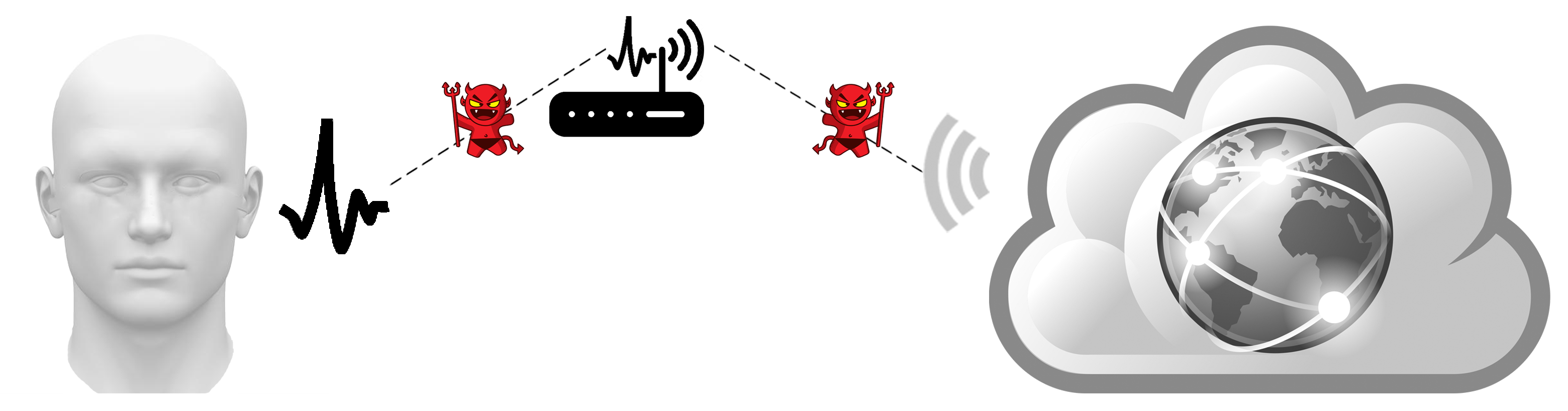}
\caption{Possible attacker locations in Configuration 2}
\label{fig:threatbrainwaves}
\end{figure}

\begin{figure}[!t]
\includegraphics[scale=0.19]{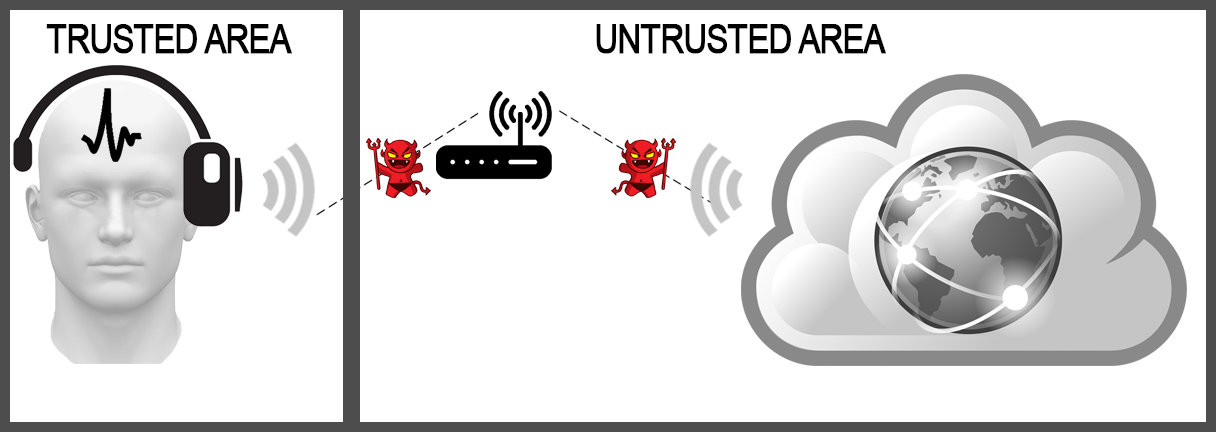}
\caption{Trusted area in Configuration 1}
\label{fig:trustthreat}
\end{figure}


\item \emph{Where is the attacker?} For instance, can the attacker attack all communication channels in the two configurations as in Fig.s~\ref{fig:threatdevice} and~\ref{fig:threatbrainwaves}? Or can we assume that the system contains a trusted network area? 
For example, Fig.~\ref{fig:trustthreat} assumes that the short-range channel between brain and device cannot be attacked, perhaps supposing that the device itself is able to provide a kind of shield creating some ``noise'' that isolates the human brain and prevents remote reading (and writing) of brainwaves, like noise-cancellation headphones do with the urban noise. Another approach could be to establish some kind of ``encryption'' between brain and device,  mapping device signals to a specific person's individual brainwaves. Alternatively, a more radical way would be to ``implant'' the device preventing possible substitutions with tampered devices. Other approaches could be possible. This situation is similar to the assumptions that are currently often made when reasoning about the security of complex security protocols (such as those built by composing subprotocols)~\cite{AlmousaMMV15,ModersheimV14,ModersheimV09} or of cyber-physical systems~\cite{LanotteMMV17}, where the attacker can only tamper with some, but not all, channels and devices. We thus expect that these recent works will be particularly useful.\fix{Luca}{I will add more here}

\item \emph{What is the power of the attacker?} What are his computational resources? Does he possess a certain amount of computation time to devote to his attack? Does he possess, or control, devices that allow him to access the different channels and the messages sent on them?
Or perhaps should we assume that the attacker can inject some malicious code in the device or the router? In that way, he could not only do harm to the system or even spoof a router to gain access to the human brain, but perhaps also physical harm to the human, by tampering with the device that has direct access to the brain. The attacker could also spoof another human to gain access to a router. We will return to this when we discuss security properties in the next subsection. In fact, we must also answer the question: \emph{What is the attacker trying to achieve?} What can he do on the different channels? Read, replace, modify, intercept messages and perhaps even brainwaves? To that end, we need to consider the security properties that the system is trying to achieve.
\end{itemize}

\subsection{Properties}

Let us now discuss the main security properties that we could ask the Internet of Neurons to guarantee. Note that although we focus on the traditional security properties, it is obvious that the categorical imperative of the Internet of Neurons is actually the \emph{safety} of the human being, i.e., 
\begin{center}
no harm should occur to the (brain of the) human being.
\end{center}
The Internet is already putting human safety at risk in several ways nowadays~\cite{kiley2002does,kuss2013internet,lee2013substance}, but in the Internet of Neurons failure to guarantee one or more security properties (e.g., consequences of the Internet ``tampering'' directly the human brain) might actually expose, directly or indirectly, humans 
to novel and much more dangerous risks.


\subsubsection{Privacy, Confidentiality and Authentication}

\emph{Information privacy} (a.k.a.~\emph{data privacy}) is the relationship between the collection and dissemination of data, technology, the public expectation of privacy, and the legal and political issues surrounding them. \emph{Internet privacy} is a subset of information privacy that concerns the storing, repurposing, provision to third parties, and displaying of information pertaining to oneself by means of the Internet. 
In the Internet of Neurons, our ``persona'' is using one of the most private information we have: our thoughts, represented by brainwaves.

Thoughts and emotions are intrinsically and intricately related. In psychology, emotions are described as unconscious feelings that are the result of mostly unconscious thoughts~\cite{pettinelli2011psychology}. A number of works have been published on how to extract human emotions from brainwaves using electroencephalography (EEG)~\cite{ismail2016human, kao2015brainwaves, lahane2015approach, chai2010classification}. What would happen if the attacker were able to extract our emotions from the brainwaves that we are sending in the Internet of Neurons? How can we protect them from being stolen? 

In Configuration 1 (as shown in Fig.~\ref{fig:threatdevice}), the attacker could intercept the brainwaves received by the device before  they are coded and transmitted to the router and then the network. A study carried out in 2011 demonstrated technologies able to reconstruct images from brainwaves~\cite{berkley}, so that, also thanks to some spoofing techniques, the attacker could intercept our communication, reverse it into brainwaves and thus obtain the raw data of our thoughts, even in their binary version. This hypothesis becomes even stronger if we consider a device-less configuration (as shown in Fig.~\ref{fig:threatbrainwaves}) where there is no encoding of brainwaves and they are broadcast over the air to the Internet. This is reminiscent of the attacks that can be carried out by eavesdropping from a distance on the sound emanated by different keyboard keys~\cite{Asonov04,Zhuang2009} or by eavesdropping from a distance on the data that is displayed on a computer screen~\cite{BackesDU08}. In these two kinds of attacks, the attacker learns how to recognize and reconstruct the sound or image generated. We expect that advances in machine learning, coupled with those in neuroscience and brainwave-data translation, will make brainwave eavesdropping and reconstruction possible with affordable attacking devices.

Another major issue concerns \textit{location privacy}. Several indoor and outdoor location techniques can be used to trace our position~\cite{gpsproc, werner2014basic}, which can have positive or negative consequences. For instance, in 2011, the Chinese government announced that it would track people's movements through their cell phones for better traffic control~\cite{guardianchina,landwehr2012privacy}, while a study of the Haitian population after the 2010 earthquake showed that similar tracking is extremely useful in informing where people are and where relief aid should go~\cite{bengtsson2011improved}. The Internet of Neurons won't be exempt from mass surveillance issues, allowing attackers, including governments or Internet providers, to violate the users' location privacy.


We could assume that every brainwave-data device will have a unique identifier like most of the devices have, such as a uuid~\cite{leach2005universally} or a global identifier that is created when the device accesses some services~\cite{jones2012creating}. Tracking these identifiers will be possible, e.g., along the lines of~\cite{koneru1999tracking}. 
Removing the device (and its identifier) as is done in Configuration 2, will help mitigate these problems, but still it won't guarantee location privacy. 
Recent studies~\cite{armstrong2015brainprint, kumari2014brainwave, ruiz2016cerebre} have namely shown that it is possible to create brainwave patterns to identify users, and thus use \emph{brainprinting} as a biometric \emph{authentication} factor.\footnote{Other studies~\cite{ThorpeOS05,MerrillCC17} have investigated \emph{pass-thought authentication}, which allows users to submit both a knowledge factor (i.e., a secret thought) and an inherence factor (i.e., the unique way that thought is expressed) in a single step, by performing a single mental task.}\fix{Luca}{We should add more here, including the original paper on passthought}
In both of the configurations that we considered, with or without a device, the attacker could then track a specific user relying just on her brainprint. To that end, the attacker would, of course, need to know the user's brainprint, but, mimicking how authentication is done today, we could imagine a sort of brainprint certificate issued by a certification authority of a public-brainprint infrastructure\footnote{The process behind the brainwave authentication methods that have been proposed requires the registration of a brainwave pattern: a sequence of images or a sequence of words are shown to a user and her brainwaves are stored as her brainprint. This process has to be done in exactly the same way for each user in order to obtain an impartial brainprint. Through this brainprint, an authentication system is able to recognize a user and then, if desired, to authenticate her requests. Note that in Configuration 1 we will also need to authenticate, and protect, the pairing of brain and device.}, or we could simply consider the Internet or the Internet provider as the attacker able to track the movement of its users.

In this case, in order to attempt to achieve location privacy, users should try to change their brainprint. One way to alter one's thought pattern would be to learn to think differently than usual, e.g., thinking ``happy thoughts'' that obfuscate the normal pattern. This sounds a bit ``mystical'', but maybe one could indeed learn to confuse one's own brainwaves while still functioning normally as a human being. Alcohol and drugs might help here (although it might then be difficult to remember one's password~\cite{drunk}) or also physical exercise, workout, fatigue, hunger and stress, which all have been shown to alter one's EEG~\cite{chuangpassthoughts}.

Another solution for privacy and location privacy, as well as for confidentiality, would be to encrypt. However, while we could use standard encryption algorithms (such as RSA, Triple DES or AES) to encrypt the wireless communication from device to router and from router to Internet, it is at best unclear how to encrypt the actual brainwaves, which are transmitted from brain to device in Configuration 1 or broadcast over the air in Configuration 2. But maybe one day somebody will devise an algorithm that allows humans to carry out mental encryption 
much in the same way as one can learn how to carry out mental calculations.


The device of Configuration~1 could raise other privacy questions. For instance, it could determine health-related issues while it is reading the user's brainwaves and provide, or sell, such information to health-insurance companies or the government. Could it also determine the user's emotions and thoughts? Will the user trust the device? How could we protect information that we know (e.g., passwords or other confidential data) from being read and distributed by the device? One could, similar to ``happy thoughts'' above, try to suppress one's thoughts about such confidential information when wearing the device, but this will be difficult if not impossible.\footnote{This is reminiscent of the \emph{paradox of thought suppression}~\cite{Wegner94}, which originates from a challenge that Fyodor Dostoevsky posed in his 1863 essay ``Winter Notes on Summer Impressions'':
\emph{Try to pose for yourself this task: not to think of a polar bear, and you will see that the cursed thing will come to mind every minute}.} Or one could learn to store some thoughts in \emph{private mental drawers}, like some mentalists are (supposedly) able to do.
In any case, to ensure that users will trust the device, it will at the very least be necessary to carry out a strict procedure of testing and certification of the device before it is deployed. Similar comments apply also for Configuration~2, but referring to the router rather than to the wearable device.



\subsubsection{Integrity}
What does integrity mean in the Internet of Neurons? How can we protect thoughts and brainwaves? The attacker will attempt to tamper with all communication channels, the digital and the mental ones. In the case of digital channels (from device to router or from router to the network), we will likely be able to use integrity-preserving solutions similar to the ones that are available now (cryptographic checksums, hash functions, message authentication codes, digital signatures, and so on).\footnote{In the case of analog channels and signals (from the device to the brain or from the brain to the router), integrity of analog brainwaves could be evaluated in the same way in which we recognize a friend's voice: first by recognition of familiar analog speech sounds, then by recognition of familiar linguistic patterns, and eventually by recognition of familiar behavioral cues and, if needed, through private shared history.}

There is of course also the question of the integrity of the human mind itself, i.e., protecting the brain from ``malicious brainwaves'' generated from malicious data from the network. In this case, we will need techniques for mental firewalls, input sanitization, sandboxing or Chinese-walling, thereby ensuring the security of the information contained in the other parts of the brain.

\subsubsection{Availability}

Besides for malfunctioning of the device and the router, and of jamming of the wireless signals, availability in the Internet of Neurons can be threatened by a \emph{Distributed Denial of Service (DDoS)} attack when the brain is overwhelmed by the amount of incoming information, thus putting the human at risk. Filtering mechanisms will be necessary to control the flow of data. 

On the other hand, the Internet of Neurons will enable opportunities that are unthinkable now. For instance, studies about sleep-learning~\cite{simon1955learning, rudoy2009strengthening, antony2012cued, arzi2012humans} have shown that our mind is able to learn if it is stimulated during the night under certain conditions. The Internet of Neurons would enable us to learn while we are sleeping thanks to the direct connection of our brain to the Internet. Actually, we could be learning in every waking moment, committing part of our brain to learning and leaving the remaining part untouched for everyday operations, i.e., for our brain's normal daily activity. We could even commit part of our brain as a CPU, e.g., for mining and other cryptographic calculations, as we have imagined in~\cite{MMM}.

\subsubsection{Anonymity}

One way to achieve at least some degree of anonymity in today's Internet is to use an anonymizing service (such as Mixes, I2P or TOR) that addresses the issue of IP tracking~\cite{zantout2011i2p,reed1998anonymous} by encrypting packets within multiple layers of encryption. Anonymity is achievable because, as the packet follows a predetermined route through the anonymizing network, each router sees the previous router as the origin and the next router as the destination, and no router knows both the true origin and the true destination of the packet. 

In Configuration 1 of the Internet of Neurons, some of the nodes of the network are actually other users with their devices, whereas other nodes are classic nodes like routers, computers and so on. In this case, the device could negotiate a preemptive path passing through a number of other devices creating a sort of onion routing. However, this kind of solution might not be applicable in Configuration 2 because it is unclear who would actually negotiate a route and apply multiple layers of encryption, unless we assume that brains are able to connect directly with each other, which is something that we will discuss in a bit more detail as we draw our conclusions.


\section{Conclusions}
\label{sec:concl}


The premise of this paper is that in 2038, in 20 years from now, the human brain will be at the center of a new Internet paradigm that we call Internet of Neurons. Some parts of our paper are deliberately science fiction (almost in the style of the Black Mirror TV series or other futuristic series and movies), but actually, as we have shown by means of the many ongoing works that we discussed, the seeds of the 
Internet of Neurons are already present in several of the technologies that are being used today or are under development. The opportunities will be prodigious, but repercussions for privacy, security and trust will be enormous and, frankly, tremendously scary.
We have thus tried to dissect some of those challenges that researchers will have to face once this is all real (and trust us, it will become real in one form or the other). However, we have only skimmed the surface. 

\begin{figure}[t]
\includegraphics[scale=0.075]{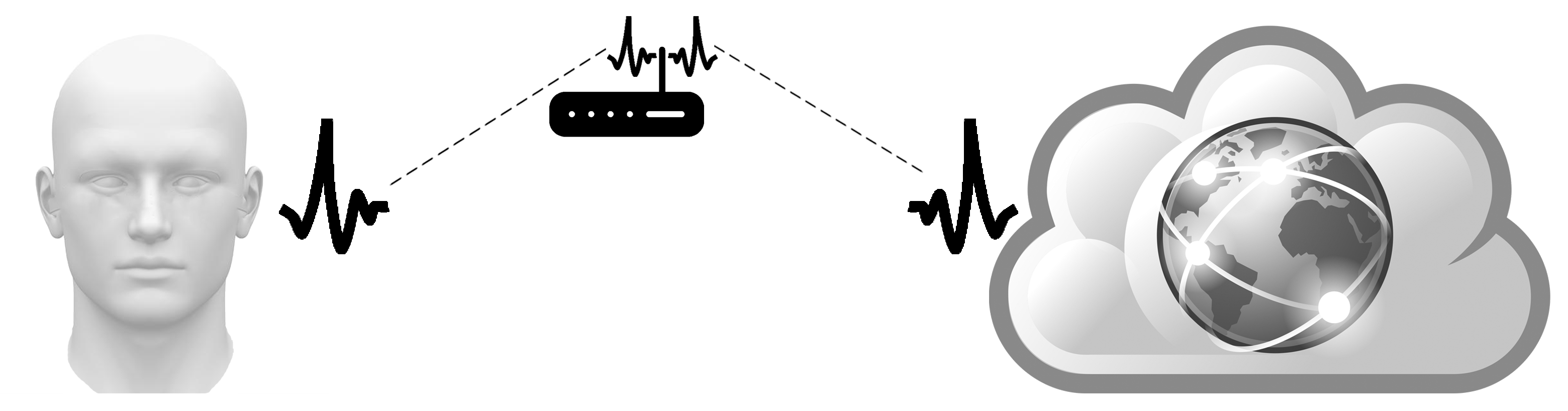}
\caption{Bi-directional brain-Internet connection by means of brainwaves}
\label{fig:brainwavesinternet}
\end{figure}

\begin{figure}[t]
\includegraphics[scale=0.075]{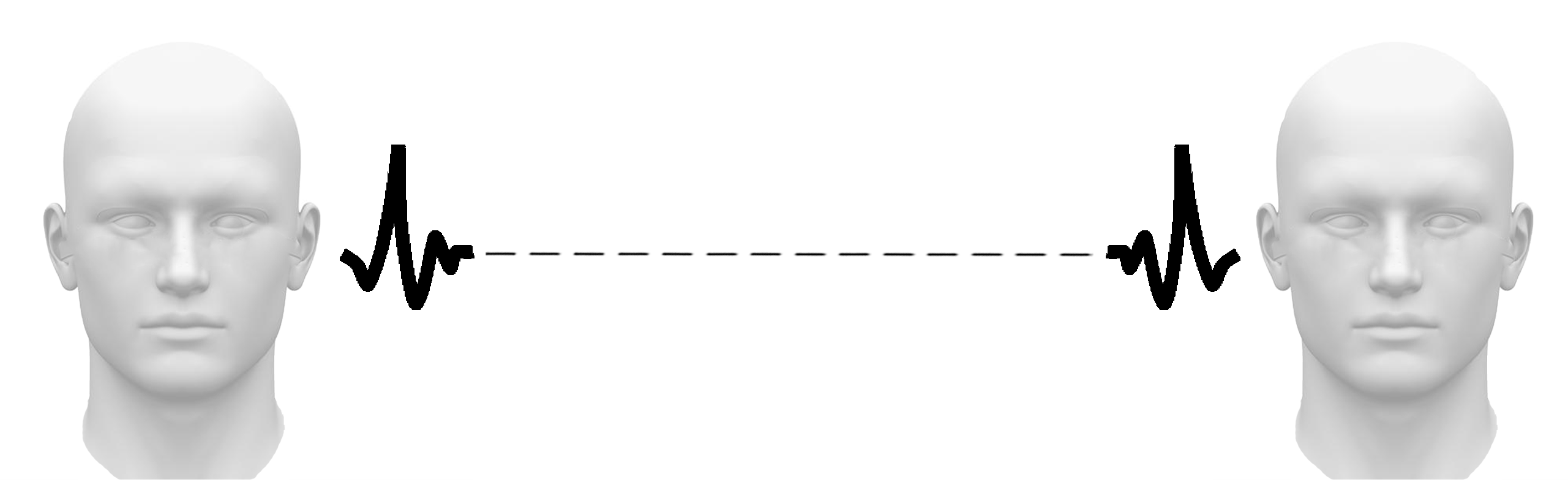}
\caption{Bi-directional brain-brain direct connection} 
\label{fig:brain-brain}
\end{figure}

More work is needed to fully understand and reason about system and threat models and security properties, specifying the ones we discussed above in more detail but also considering other properties that could be relevant for the Internet of Neurons.
Moreover, we have made the quite strong assumption that brainwaves will need to be translated to data (and vice versa) as the Internet will still transmit packets. But by, say, 2050, it could well be that the network will follow a radically different model, perhaps thanks to advances in quantum computing or in ``brainwave computing'' (a discipline that we just invented), allowing the network to directly process brainwaves as shown in Fig.~\ref{fig:brainwavesinternet}. But why stop here? If brainwave transmission protocols are possible, then it means that the network is able to read the brainwaves that a brain is emanating, but also that the brain is able to receive brainwaves in input. How long will it then take before we find a way for brains to connect not only to the network but also to each other? Some research in this direction is already ongoing~\cite{grau2014conscious, 0111332} and the ultimate Internet of Neurons might then simply be based on direct brain-brain connections as the one in Fig.~\ref{fig:brain-brain}. 

Finally, there is an elephant in the room that we have not addressed in this paper. In addition to technological and neurological questions, some of which we discussed above, there are a huge number of economical, political and ethical issues that we don't really feel competent to address, but that will have to be tackled before we open our mind to the Internet. Who will pay for the Internet of Neurons? Will all citizens be taxed? Will governments or perhaps corporations provide it for free? Given that nothing is actually free, what will they want in return? In the wake of the recent scandals on data collection (such as the Facebook--Cambridge Analytica data scandal that involved the collection of personally identifiable information of up to 87 million Facebook users), we are skeptical that the Internet of Neurons will be exempt from massive personal data collection and mining, possibly opening up the possibility for big-brother scenarios in which citizens are always observed and tracked in order to control and influence their thoughts, opinions, votes, in brief, their whole life.

\end{document}